\documentstyle[aps, prl, epsf, multicol]{revtex}

\begin{document}
\title{Phase diagram of an asymmetric spin ladder} 
\author{Shu Chen, H. B\"{u}ttner, J. Voit}
\address{
Theoretische Physik 1, Universit\"{a}t Bayreuth, D-95440 Bayreuth, Germany}
\date{\today}
\maketitle

\begin{abstract}
We investigate an asymmetric zig-zag
spin ladder with different exchange integrals on both legs
using bosonization and renormalization group. 
When the leg exchange integrals and frustration both are sufficiently small, 
renormalization group analysis shows that
the Heisenberg critical point flows to an
intermediate-coupling fixed point with gapless excitations and a 
vanishing spin velocity. When they are large, a spin gap opens
and a dimer liquid is realized. Here, we find 
a continuous manifold of Hamiltonians with  dimer product ground states,
interpolating between the Majumdar-Ghosh and sawtooth spin-chain model. 


\end{abstract}

\begin{multicols}{2}
\narrowtext


The interplay of geometric frustration and quantum fluctuations,
and eventually broken translational invariance,  in 
low-dimensional spin systems gives rise to novel magnetic phases. 
Examples include: spin chains with interactions beyond nearest
neighbors, spin ladders, triangular and Kagom\'{e} systems\cite{moessner}.
Spin ladders in particular, have attracted much interest and several
model problems are well understood\cite{Dagotto}, and highlight the
role played by frustration. Spin-isotropic two-leg ladders with railroad 
geometry quite generally lead to a singlet ground state separated by
a finite excitation gap from the first triplet states\cite{Dagotto}. Zig-zag
ladders are more strongly frustrated and, depending on
the ratio of the leg to rung exchange integrals, 
may have gapless spin liquid
ground states or gapped dimer states\cite{Haldane}. 
Compounds like $Cu_2(C_5H_{12}N_2)_2 Cl_4$ and $CuGeO_3$ have been suggested 
to be described as railroad and zig-zag ladders, respectively \cite{Castilla}.

Little work has been done on asymmetric spin ladders where the exchange integrals
on both legs differ. Only the extreme case where one leg of a zig-zag
ladder is missing entirely (sawtooth or $\Delta$-chain) has been 
solved\cite{Sen,Nakamura}. The ground state of this model is a product of 
nearest-neighbor singlets and there is a spin gap, as in the Majumdar-Ghosh 
model\cite{M-G}. 
The properties of the excitations in both models are different, however.
Sawtooth chains may describe the material $YCuO_{2.5}$ \cite{Sen,Nakamura}.
 Here, we perform a 
systematic study of asymmetric zig-zag spin ladders. Important questions
concern possible new phases generated by the leg-asymmetry, their excitation 
spectra and their physical properties as well quantitative
changes brought about by leg-asymmetry to the excitations in the more usual 
phases found on zig-zag ladders (spin liquid, N\'{e}el and dimer states). 

\begin{figure}[t]
\centerline{\epsfysize=2.5cm \epsffile{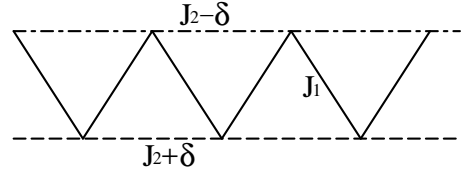}}
\caption{The asymmetric zig-zag spin ladder.}
\label{mod_fig}
\end{figure}

We consider a Heisenberg model on the structure shown in Figure \ref{mod_fig},
and represent this as a chain with an alternating next-nearest-neighbor
(NNN) exchange
\begin{equation}
\label{eq1}
H=\sum_{l} \left\{ J_{1}{\bf S}_{l} \cdot {\bf S}_{l+1}+
\left[ J_{2}+(-1)^{l}\delta \right]
{\bf S}_{l} \cdot {\bf S}_{l+2} \right\} \; , 
\end{equation}
where $J_{1}\equiv 1$ and $J_{2}\pm \delta$ are the nearest-neighbor (NN) 
and alternating NNN coupling constants, respectively. 
$\delta=0$ is the ordinary zig-zag ladder or frustrated spin 
chain\cite{Haldane}, and fixing in addition 
$J_{2}={\frac{1}{2}}J_{1}$ gives the exactly solved Majumdar-Ghosh (M-G)
model \cite{M-G}. 
The sawtooth chain has $\delta =J_{2}={\frac{1}{2}}J_{1}$\cite{Sen,Nakamura}. 

Following the general approach to map the quantum spin model to a continuum 
field theory\cite{Affleck} and using the standard dictionary of Abelian 
bosonization\cite{Voit1}, we obtain an effective boson Hamiltonian $H = 
H_0 + H_1$ with 
\begin{eqnarray}
H_{0} & = & \int dx\frac{u}{2\pi }\left[ K(\pi \Pi )^{2}+\frac{1}{K}(\partial
_{x}\Phi )^{2}\right] \; ,  \label{h0} \\
H_1 & = & \int dx \left[ 
{\frac{g_{3}}{2(\pi a)^{2}}}\cos 4\Phi+{\frac{g_{1}}{\pi ^{2}a}}
\left( \partial_{x} {\Phi} \right) \cos 2\Phi
\right] \; . \label{h1}
\end{eqnarray}
$\Phi(x)$ is a bosonic phase field and $\Pi(x)$ its canonically conjugate
momentum. 
$u$ and $K$ are the effective spin velocity and coupling constants, including
the effects of marginal interactions.
$a$ is a short-distance cutoff, $g_{3} \propto 1- J_{2}/J_{2c}$ is the 
Umklapp-scattering amplitude and $g_{1} \propto \delta$ is the amplitude 
of the alternating NNN field. In this paper, we only consider the case 
$J_1/2 \geq J_{2} > \delta$, while,  for $J_{2}>0.5J_{1}$, quite different  
field theory treatments are required\cite{Allen2,White}. 

For $\delta = 0$, the model (\ref{eq1}) is well understood
\cite{Haldane}. Its elementary excitations are spinons which are gapless
for $J_2 < J_{2c}=0.2412$\cite{Castilla} when frustration is irrelevant
and the ground state is unique, 
or gapped for $J_2 > J_{2c}$ when frustration is relevant
and the ground state is doubly degenerate. In the field
theory (\ref{h0}), (\ref{h1}), this translates into the $g_3$ term being
either marginally irrelevant, leading to the weak-coupling
Heisenberg fixed point ($K^{(H)} =1/2, \; g_3^{(H)}=0$)  \cite{Affleck},
or relevant, yielding a strong-coupling dimer state. When the SU(2) spin
symmetry is broken, a N\'{e}el state or an easy-plane spin liquid may form.
Introducing dimerization, i.e. an alternating $J_1$, produces an
effective confining potential between the spinons. The elementary
excitations become a spin triplet and a spin singlet above the unique ground
state. In the language of field theory, the external dimerization corresponds
to a relevant term $\sin 2\Phi$ which lifts the double degeneracy of the dimer
ground state\cite{Affleck2}.

Qualitative results on the influence of the new interaction ($g_1$) can
be obtained from physical considerations alone. Firstly, the
scaling dimensions of the Umklapp and nearest neighbor alternation terms
$g_3$ and $g_1$ are 
\begin{equation} d_{g_{3}}=4K\; ,~~d_{g_{1}}=K+1 \; .
\label{scaldim}
\end{equation}
At the Heisenberg fixed point, $g_{3}$ is
marginal  with $d_{g_{3}}=2$, while the $g_{1}$-term with $d_{g_{1}}= 3/2$ 
is relevant. 
We conclude that $g_1$ destabilizes the isotropic Heisenberg
fixed point and the spin liquid ground state. On the other hand, there is
no standard strong coupling theory for the $g_1$-term. Usually (e.g. 
$g_3 \rightarrow \pm \infty$), the boson field $\Phi(x)$ locks into a constant
value with small fluctuations, and an associated excitation gap. 
Such a phase locking, however, is forbidden by the $\partial_x \Phi$-prefactor
to the cos$(2 \Phi)$-term in $H_1$. 
Secondly, the $g_1$-term, induced by the alternating NNN
interaction, does not confine the spinons and plays a role very different
from the external dimerization, as can be checked in the dimer state by
comparing the sawtooth chain\cite{Sen,Nakamura} with the M-G model\cite{M-G}.
Moreover, from differences in the size of the
spin gaps in these two models, it follows
that the $g_1$-term quite generally counteracts the Umklapp term 
while an external NN dimerization would cooperate. 

That $g_1$ opens no spin gap  despite 
being a relevant perturbation of the Heisenberg fixed point, is also corroborated
by the absence of a magnetization plateau in our model in small magnetic 
fields\cite{Wiessner}. A necessary condition for the formation of a magnetization
plateau in the absence of modulated external fields is an alternating component
of the exchange integrals \cite{Oshikaw,Fled}. For an alternating NN exchange,
a magnetization plateau is observed in small magnetic fields, but alternating
NNN exchange, it is only observed in high fields \cite{Wiessner,Fled}. 

We now perform a perturbative renormalization group (RG) analysis by mapping
the model on a modified (by $g_1$) classical 2D XY-model\cite{chui,KT}.
Introducing the reduced variables $y_{3}=g_{3}/\pi u$ and
$y_{1}=g_{1}/\sqrt{2}\pi u$, we obtain the linearized RG equations
\begin{eqnarray}
\label{dk}
\frac{d~K}{dl}\, &=&\,-y_{3}^{2}{K}^{2}+y_{1}^{2}{K}^{4}\, \\
\label{dy3}
\frac{d~y_{3}}{dl}\, &=&\,(2-4K)y_{3}+K^{2}{y_{1}}^{2}\, \\
\label{dy1}
\frac{d~y_{1}}{dl}\, &=&\,(1-K)y_{1}-4K^{2}y_{1}y_{3}\, \\
\label{du}
\frac{d~u}{dl}\, &=&\,-{\frac{1}{2}}uy_{1}^{2}(1+K){K}^{2}.
\end{eqnarray}
under a change of length scale $a \rightarrow a e^{dl}$. 
These equations, and our solutions to be discussed below, can also accommodate
anisotropy in the NN and NNN exchange integrals. The RG equation
for the spin velocity $u$ is a consequence of the anisotropy of the 
$g_1$-interaction in the classical 2D XY-model, i.e. its non-retarded but 
non-local character in the quantum field theory (\ref{h1}),
and has been discussed before in the 1D electron-phonon problem\cite{Voit2}.
For $y_1=0$, one finds the three
generic phases discussed earlier\cite{Haldane}: (i) a weak-coupling 
spin liquid phase ($g_3^{\star}=0, \; K^{\star} \geq 1/2$) terminating at the
isotropic Heisenberg fixed point $K^{(H)}=1/2$; (ii) a strong coupling N\'{e}el
phase $g_{3}^{\star} \rightarrow \infty$; (iii) a strong coupling dimer phase
$g_3^{\star} \rightarrow -\infty$. Moreover, $u$ is not renormalized
for $y_1 = 0$. Spin-rotation invariant 
models scale along the separatrix between N\'{e}el and spin liquid phases, or
along its continuation into the dimer regime. 

\begin{figure}
\centerline{\epsfysize=6.0cm \epsffile{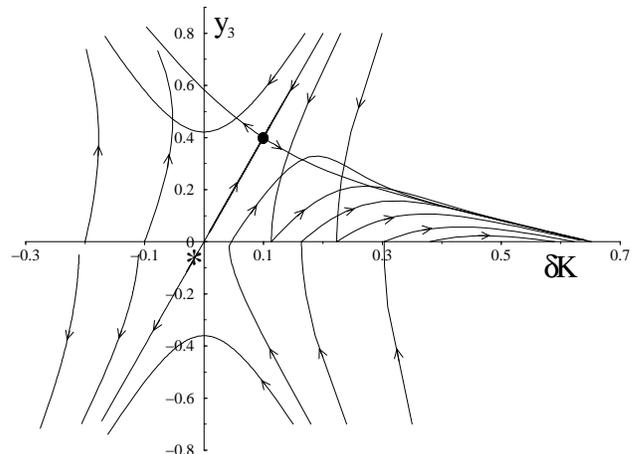}}
\caption{The scaling trajectories for $y_1( l =0) =0.001$ 
projected on the $y_3 - \delta K$ plane. $\delta K = K-1/2$, and the dot 
locates the new intermediate coupling fixed point. The N\'{e}el state is
realized in the upper left, the dimer state in the lower left, and the 
spin liquid in the right part of the figure. The star locates the boundary
between flows to the new fixed point, and into the dimer regime.}
\label{rgroup}
\end{figure}

Taking $y_1$, the most relevant perturbation, finite, changes this simple 
picture. Quite generally, the corrections to the RG flow of $K$ and $g_3$
are \em positive, \rm while those from $g_3$ are negative. In the spin liquid
phase, close to the isotropic Heisenberg fixed point, the 
alternating NNN exchange will \em increase \rm both $K$ and
$y_3$, and therefore \em reverse the direction \rm of the RG flow, compared 
to the $y_1=0$ situation. The sign of this effect agrees with the spin
gap reduction in the dimer phase when going from the M-G to the sawtooth
model. The magnitude of the effect is a direct consequence of the
scaling dimension $d_{g_1}$. Figure \ref{rgroup} shows a family of solutions
of the RG equations, projected on the $y_3-K$-plane,
which have been linearized for the purpose of the figure
to accurately control spin-rotation invariance (cf. below).
The flow reversal is seen in much of the first quadrant. When 
$K$ is increased sufficiently by $y_1$, this perturbation becomes irrelevant,
however, and the RG flow bends back to a weak coupling fixed line ($K^{\star}
\geq 1, y_3^{\star} = 0, y_1^{\star}=0$), describing a spin liquid with an
increased easy-plane anisotropy. Its spin velocity $u$ is decreased but remains
finite.

More interesting is the spin-rotation invariant line ($y_3 = 4K-2$ in the 
linearized RG). While the initial flow towards the Heisenberg fixed point
also reverses its direction under the influence of $y_1$, it reaches a
new finite-coupling fixed point
$ (K^{\star} \approx 0.578, y_{3}^{\star}  \approx 0.315, y_{1}^{\star}
 \approx \pm 0.544)$,
resp.\ $ (K^{\star} = 0.6, y_{3}^{\star} = 0.4, y_{1}^{\star} = \pm 0.8)$ 
in the linearized RG. This fixed point is attractive for spin-rotation 
invariant  systems, and repulsive otherwise with a flow into the spin liquid
 or N\'{e}el phases for easy plane (easy axis) anisotropy. 
At the fixed point, the renormalized spin velocity vanishes, 
$u^{\star} = 0$. This conclusion follows from Eq.\ 
(\ref{du}) and is independent of the 
exact value of the fixed point as long as it is located on the RG separatrix 
with $1/2 <  K < \infty$.
Consequently, conformal invariance is broken, and we do not expect power-law 
decay of the correlation functions. Thermodynamic properties will be 
different from a Heisenberg chain: the specific heat will show nonlinear 
temperature dependence, and the magnetic susceptibility will depend on 
temperature.  This state is 
significantly different from the Luttinger-type 
spin fluid state of the Heisenberg model. The fixed-point Hamiltonian can be
rewritten as $H^{\star} = u^{\star} {\cal H}(K^{\star}, g_1^{\star},
g_3^{\star})$ where ${\cal H}$ is independent of $u^{\star}$. It becomes
trivial, $H^{\star} = 0$, at the fixed point because $u^{\star} = 0$. 
We interpret this as our spins effectively decoupling at 
the lowest energy scales, i.e. a kind of asymptotic freedom in this
spin-rotation invariant ladder. 

From standard arguments, 
we find that the elementary excitations in 
the fixed point, spinon and antispinon, are still gapless. 
One can also provide a variational argument, based on an effective 
Hamiltonian $H_1$, Eq.\ (\ref{h1}). Minimizing its energy 
 directly produces pairwise kink and antikink solutions corresponding to 
the elementary excitations with
energies proportional to $g_3$ resp.\ $g_1$. Including the quantum
fluctuations from $H_0$ is expected to delocalize these kinks and reduce their
energies further. Using now our fixed point properties
$g_{1,3} = y_{1,3}^{\star} u^{\star}$, $u^{\star} \rightarrow 0$ makes the
excitation gap vanish at the fixed point. 
Also, the numerical results of Wiessner et
al.,\cite{Wiessner}, indicate a paramagnetic susceptibility.

A natural question arises here: Is this intermediate fixed point stable 
against higher order perturbations? Our answer is positive although
we do not attach any particular significance to the 
numerical values of the coupling constants, which may change as
higher-order corrections are included. Its location on the
RG separatrix between the N\'{e}el state and the easy-plane spin liquid
is protected by spin-rotation
invariance and it can be pushed neither back to the Heisenberg fixed point,
nor to strong coupling, $K \rightarrow \infty$, by higher-order operators.
The first option is inconsistent with the scaling dimensions, the second
one would correspond, in the fermion
language, to long-range pairing order which is excluded in 1D. 
Scaling into the dimer regime is 
inconsistent with the absence of a magnetization plateau\cite{Wiessner}.

Care should be taken when comparing these predictions to results, e.g., 
from exact numerical diagonalization on small clusters (size $N$). 
The finite system size
will stop the RG flow at a scale $l_N = \ln N$, and the fixed point ($l 
\rightarrow \infty$) is \em not \rm reached. However, Eqs.\ (\ref{dk})--(\ref{du})
predict a spin velocity decreased significantly by the alternating component
of the NNN exchange, with an unusual size dependence which can be evaluated
by integrating the RG equations up to $l_N$. Due to the scale dependence of 
the renormalization, the finite-size spectrum likely exhibits significant 
nonlinear corrections which, again, are size-dependent. The velocity 
renormalization also suggests a scale-dependence of the magnetic susceptibility.
A direct solution of our model in a magnetic field is rather involved, however,
because higher-order terms lead to a field-induced generation of new, relevant
operators. This will be reported later. 

Of course, when $J_2$ increases beyond a critical value $J_{2c}(\delta)$ now
depending on the exchange alternation, the RG flows to a strong coupling fixed 
point, which corresponds to the quantum 
dimer phase. For $y_1 = 0.001$, this critical point is indicated in 
$(K,y_3)$-coordinates in Figure 2 by a star. For small 
$\delta$ and $J_2 > J_{2c}(\delta)$, 
our RG equations show that the spin gap is decreased by increasing 
$\delta$, but the system will remain in the universality class of the dimer
liquid. 

However, we do not expect 
that the field theory description is very precise for $J_{2}$ near the M-G 
point $J_{2}=0.5J_{1}$. The main reason is that the correlation length,
which is proportional to the inverse of the energy gap, 
decreases quickly when $J_{2}$ increases much beyond $J_{2c}(\delta)$. 
To access this limit, we now discuss the influence of NNN exchange alternation
on the ground state and excitations of the Majumdar-Ghosh model.
For the M-G model\cite{M-G}, the two linearly independent
ground states, say, left or right dimer ground state, are products 
of nearest-neighborly singlets, respectively 
\begin{equation}
\label{mggstate}
\mid \Phi _{L}\rangle =\prod_{l=odd}[l,l+1],~~\mid \Phi _{R}\rangle
=\prod_{l=even}[l,l+1],
\end{equation}
where 
$
[i,j]=([\uparrow ]_{i}[\downarrow ]_{j}-[\downarrow ]_{i}
[\uparrow]_{j})/\sqrt{2} ~~~
$
denotes the singlet combinations of spin i and j. (\ref{mggstate}) also
represents the degenerated ground state of the 
sawtooth model\cite{Sen,Nakamura}. How are these models connected?
We notice that NNN exchange alternation does not modify the product states
of nearest-neighbor singlets
\begin{equation}
H_{\delta }\mid \Phi _{L,R}\rangle = \sum
(-1)^{l}\delta {\bf S}_{l}{\bf S}_{l+2}\mid \Phi
_{L,R}\rangle =0.
\end{equation}
Furthermore, we can prove that
 introducing NNN exchange alternation into the M-G model, or 
changing it in the sawtooth chain, will not affect their ground states.
There is thus an entire manifold of Hamiltonians, parameterized by $\delta$
with $J_1 = 2 J_2$ fixed,
with doubly degenerate NN-dimer product ground states
$| \Phi_{L,R} \rangle$. 
This kind of ground state attracted much attention
recently for the experimental realization of $SrCu_2(BO_3)_2$ as
2D Shastry-Sutherland model\cite{Ueda,S-S}.

While in these models, the focus is on the ground state, 
we also can characterize the elementary excitations as 
kinks and antikinks with finite excitation energies and different 
dispersions, in the dimer state of our ladder. 
Starting from the M-G model, the kinks
can be thought of as domain walls separating the different
dimer ground state configurations. From  symmetry considerations, the kink and
antikink properties are identical in the M-G  model. With
alternating NNN interaction, the symmetry between legs is broken, 
some properties will differ between kinks and antikinks, 
in particular the dispersions. However, they still
survive as the  elementary excitations of the asymmetric spin ladder system.
The alternating NNN interaction does not serve as a confining  potential for
the kink and antikink, but changes the energy gap of the
excitations. Based on the cluster
variation method \cite{Sen,Nakamura}, our numerical estimates indicate that the gap
decreases from $0.234$ in the M-G model to $0.219$ in the sawtooth chain with 
increasing $\delta$, while the ground state remains
invariant. Particularly, for $\delta = J_{2}$, i.e. the sawtooth chain, the
kink excitation is exactly a single spin on odd site and dispersionless, while
 an antikink is still a domain wall propagating with an effective
mass\cite{Sen,Nakamura}.

The preceding results can also be understood from the point of view of the 
corresponding field model. The presence of $g_1$ term 
$\partial_{x}\Phi \cos2\Phi$ does not lift the degeneracy of the 
dimer ground state,
 which is the constant solution minimizing the energy of the Hamiltonian with 
 $g_3$ term. Near the strong coupling dimer fixed point, the $g_3$ term is 
much more 
relevant than the $g_1$ term. In this case, the soliton solutions 
of the $\cos4\Phi$ sine-Gordon equation will survive. However, the $g_1$ term,
 although it is much less relevant than the $g_3$ term, gives different masses to
the sine-Gordon kink 
and antikink solutions. This give us a rough explanation 
to the asymmetric kink and antikink excitations in the presence of alternating
NNN coupling. 
The difference in (numerically determined) gap size between the M-G and
sawtooth chains, is consistent
with our RG results on the influence of $\delta$ for smaller $J_2$, on the
spin gap magnitude. 

Spin-isotropic, asymmetric zig-zag ladders are described, in the
limit of weak frustration, by an intermediate-coupling fixed point with
gapless  excitations and a vanishing spin velocity, likely indicating a
decoupling  of the spins at low energy scales. For larger frustration, a more
usual  dimer liquid phase is realized whose spin gap decreases with increasing
  leg-asymmetry. A continuous manifold of Hamiltonians with the same  
singlet product ground state interpolates between the Majumdar-Ghosh model 
and the sawtooth spin chain.  
In addition, we find gapless spin liquid and gapped N\'{e}el 
states with easy-plane and easy-axis anisotropy. Extensions we currently 
consider include external fields and doping with charge carriers.  

We are grateful to Prof. K. H. M\"{u}tter and Dr. M. Nakamura for 
fruitful discussions. This research was supported by Deutsche 
Forschungsgemeinschaft through grants no. VO436/6-2 (Heisenberg fellowship to
J.V.) and VO436/7-1. 

\end{multicols}

\begin{thebibliography}{99}
\bibitem{moessner} R. Moessner, cond-mat/0010301.
\bibitem{Dagotto} E. Dagotto and T. M. Rice, Science, {\bf 271}, 618 (1996)
\bibitem{Haldane} F. D. M. Haldane, Phys. Rev. B. {\bf 25}, 4925 (1982); 
	{\bf 26}, 5257 (1982)
\bibitem{Castilla} G. Chaboussant, \em et al., \rm Phys. Rev. Lett. {\bf 79}, 
	925 (1997); G. Castilla, S. Chakravarty and V. J. Emery, Phys. Rev.
	Lett. {\bf 75}, 1823 (1995)
\bibitem{Sen} D. Sen, S. Shastry, R. E. Walstedt and R. Cava, Phys. Rev. B. 
	{\bf 53}, 6401 (1996)
\bibitem{Nakamura} T. Nakamura and K. Kubo, Phys. Rev. B. {\bf 53}, 6393 (1996);
	T. Nakamura and S. Takada, Phys. Rev. B. {\bf 55}, 14413 (1997)
\bibitem{M-G} C. K. Majumdar and D. K. Ghosh, J. Math. Phys. {\bf 10}, 1388
	(1969); {\bf 10}, 1399 (1969)
\bibitem{Affleck} I. Affleck, in Les Houches, Session {\bf XLIX}, Fields,
	strings and critical phenomena, Elsevier, Ney York, 1989
\bibitem{Voit1} J. Voit, Rep. Prog. Phys. {\bf 58}, 977 (1995)
\bibitem{Allen2} D. Allen and D. S\'{e}n\'{e}chal, Phys. Rev. B. {\bf 51},
	6394 (1995)
\bibitem{White} S. R. White and I. Affleck, Phys. Rev. B. {\bf 54}, 9862
	(1996)
\bibitem{Affleck2} I. Affleck, Cond-mat/9705127; 
A. M. Tsvelik, Phys. Rev. B. {\bf 45}, 486 (1992);  G. S. 
Uhrig and H. J. Schulz, Phys. Rev. B. {\bf 54}, R9624 (1996) 
\bibitem{Wiessner} R. M. Wiessner, A. Fledderjohann, K. H. M\"{u}tter,
	and M. Karbach, Eur. Phys. J. B {\bf 15}, 475 (2000)
\bibitem{Oshikaw} M. Oshikawa, M. Yamanaka, and I. Affleck, Phys. Rev. Lett. 
	{\bf 78}, 1984 (1997)
\bibitem{Fled} A. Fledderjohann, \em et al., \rm
	Phys. Rev. B. {\bf 59}, 991 (1999)
\bibitem{chui} S.-T. Chui and P. A. Lee, Phys. Rev. Lett. {\bf 35}, 325 
	(1975) 
\bibitem{KT} J. M. Kosterlitz, J. Phys. C. {\bf 7}, 1046 (1974)
\bibitem{Voit2} J. Voit and H. J. Schulz, Phys. Rev. B. {\bf 36}, 968
	(1987); Phys. Rev. B. {\bf 37}, 10068 (1988)
\bibitem{Ueda} S. Miyahara and K. Ueda, Phys. Rev. Lett. {\bf 82}, 3701  
	(1999)  
\bibitem{S-S} B. S. Shastry and B. Sutherland, Physica B {\bf 108}, 1069
(1981);
 J. Richter, N.B. Ivanov and J. Schulenburg, J.Phys.: Condens. Matter
{\bf 10}, 
3635 (1988)
\end{thebibliography}
\end{document}